\documentclass[%
jcp,
 aip,
jmp,%
 sd,%
 amsmath,amssymb,
 reprint,%
]{revtex4-1}

\usepackage{xcolor}
\usepackage{graphicx}
\usepackage{dcolumn}
\usepackage{bm}
\usepackage{hyperref}
\usepackage{multirow}
\usepackage{etoolbox}
\patchcmd{\footnotemark}{\stepcounter{footnote}}{\refstepcounter{footnote}}{}{}
\newcommand{\x}{\ensuremath{\mathbf{x}}}

\begin{document}


\title[]{Including surface ligand effects in continuum elastic models of nanocrystal vibrations}

\author{Elizabeth M. Y. Lee}
\affiliation{%
 Department of Chemical Engineering, Massachusetts Institute of Technology,
\\Cambridge, Massachusetts 02139, USA
}%
\author{A. Jolene Mork}%

\author{Adam P. Willard}
\email{awillard@mit.edu}
\affiliation{%
 Department of Chemistry, Massachusetts Institute of Technology,
\\Cambridge, Massachusetts 02139, USA
}%
\author{William A. Tisdale}
\email{tisdale@mit.edu}
\affiliation{%
 Department of Chemical Engineering, Massachusetts Institute of Technology,
\\Cambridge, Massachusetts 02139, USA
}%

\begin{abstract}
The measured low frequency vibrational energies of some quantum dots (QDs) deviate from the predictions of traditional elastic continuum models. Recent experiments have revealed that these deviations can be tuned by changing the ligands that passivate the QD surface. This observation has led to speculation that these deviations are due to a mass-loading effect of the surface ligands. In this article, we address this speculation by formulating a continuum elastic theory that includes the mass-loading effects of the surface ligands. We demonstrate that this model is capable of accurately reproducing the $l=0$ phonon energy across variety of different QD samples, including cores with different ligand identities and epitaxially grown CdSe/CdS core/shell heterostructures. We highlight that our model performs well even in the small QD regime, where traditional elastic continuum models are especially prone to failure. Furthermore, we show that our model combined with Raman measurements can be used to infer the elastic properties of surface bound ligands, such as sound velocities and elastic moduli, that are otherwise challenging.

\end{abstract}

\maketitle
\section{Introduction}
The photophysical properties of semiconductor nanocrystals, also known as quantum dots (QDs), are affected by the underlying vibrations of the atomic lattice.~\cite{Muljarov2004,Favero2007,Fernee2008a} Unlike their bulk counterparts, semiconductor nanocrystals include low frequency, confined vibrational modes associated with particle distortions, such as the symmetric or ellipsoidal stretching. The characteristics of these collective vibrational modes are known to depend on QD size, shape, composition, and heterostructure. These dependences can be understood in terms of elastic continuum models.~\cite{Saviot1996,Verma1999,Ikezawa2001,Ivanda2003,Saviot2004,Chassaing2009,Chilla2008,Cerullo1999,Fujii1996,Zhao1999,Crut2014} Recently, however, it has been observed that these low frequency vibrational modes also depend on the identity of the ligands that passivate the surface of the QDs,~\cite{Mork2016a} presenting a challenge to existing theories. In this article, we show that the effect of ligand identity on the low frequency vibrations of QDs can be understood in the context of traditional continuum elastic theories by designing the boundary conditions to mimic viscoelastic properties of the ligand shell. This theory, as we show, is able to predict a broad range of experiments where existing approaches fail.

The prevailing methods for computing the vibrational properties of QDs are based on a theoretical model in which the nanocrystal is described as a linear elastic field whose properties are assumed to be identical to that of a bulk crystal. The size and shape of this field determine the spectrum of quantized low frequency collective vibrations, or acoustic phonon modes. Lamb~\cite{Lamb1881} first formalized this model by applying the Cauchy-Navier equation,~\cite{Eringen1975}
\begin{equation}
v_l^2 \nabla(\nabla\cdot \mathbf{u}) - v_t^2 \nabla \times (\nabla \times \mathbf{u}) =\frac{\partial^2{\mathbf{u}}}{\partial t^2},
\label{eq:1}
\end{equation}
to a perfectly spherical elastic particle, where $\mathbf{u}$ is the displacement field, and $v_l$ and $v_t$ are longitudinal and transversal sound velocities, respectively. The solution to this equation, expressed in terms of eigenmodes, is $\mathbf{u}(\mathbf{r},t) =\sum_{i} \mathbf{u}_i(\mathbf{r}) \exp{(-i\omega_i t)}$, where $\omega_i$ describes the corresponding frequency of the $i$-th eigenmode. Lamb's model predicts that frequencies of the lowest energy vibrational modes, \textit{e.g.}, radial breathing mode and ellipsoidal mode, scale inversely with QD radius, $R$.~\cite{Lamb1881} This scaling relationship has enabled the detailed assignment of Raman spectra for a wide range of different nanoparticles.~\cite{Saviot1996,Montagna1995,Ivanda2003, Ikezawa2001, Zhao1999,Fujii1996,Saviot2004} 

The proportionality constant associated with this $1/R$ scaling relationship depends on the choice of boundary conditions. Limiting cases include free boundary conditions (as in Lamb's original model), where the stress, $\bm{\sigma}$, is equal to zero at the surface, and rigid boundary conditions, where the displacement, $\mathbf{u}$, is equal to zero at the surface. In some cases, such as modeling QDs embedded in a glass matrix, the boundary conditions are treated as a dissipative bath.~\cite{Saviot2004} In the absence of a well controlled series of experiments, however, the appropriate choice of boundary condition has remained ambiguous.

Experiments have shown that in some cases, the scaling of low frequency vibrational modes deviates from the $1/R$ prediction of elastic continuum models.~\cite{Cerullo1999,Huxter2009, Mork2016a, Mork2016b} These deviations were found to be more pronounced for smaller quantum dots ($R\lesssim2$ nm), which has led to speculations about the origin of these deviations. Possible interpretations include ligands affecting the boundary condition of QD surface~\cite{Cerullo1999} and structural relaxation occurring at the surface that may lead to a size-dependent elastic moduli.~\cite{Huxter2009} Recent ligand substitution experiments on both spherical QDs and 2D nanoplatelets suggest, however, that mass loading of the nanocrystal surface is the dominant cause of these observed deviations.~\cite{Mork2016a,Girard2016} 

Here, we show that traditional elastic continuum models can be modified to include these mass loading effects. By doing so, we successfully describe small-$R$ deviations in the scaling of low frequency vibrational modes, along with their dependence on changes in ligand identity. Furthermore, we show that this model is capable of accurately predicting the vibrational resonances of more complex spherical heterostructures, including core/shell QDs.

In the following section, we describe a general framework for including the boundary effect of ligands in elastic continuum models. We then compare the results of this model to a series of low-frequency Raman scattering experiments carried out on QDs with varying core size and ligand identity. 

\section{Continuum Linear Elasticity Theory} 
\begin{figure}[h!] 
\centering
\includegraphics{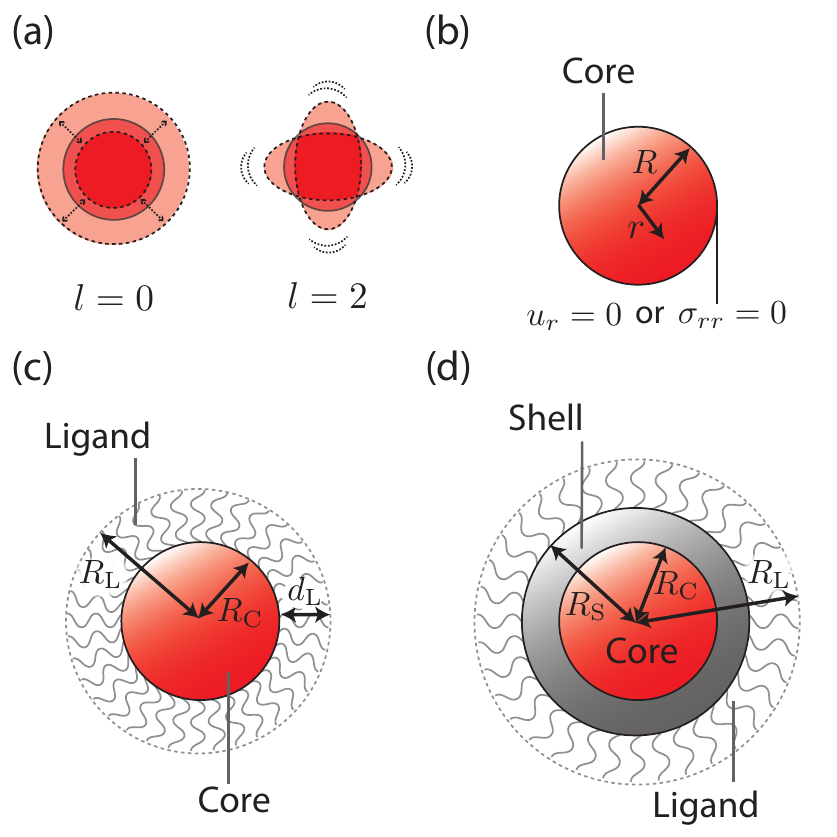}
\caption{Modeling acoustic phonon energies of colloidal quantum dots. (a) Schematic illustrations of the Raman-active acoustic eigenmodes of elastic spheres ($l=0$ and $l=2$ spheroidal vibrations). Diagrams of (b) the core-only, (c) the core-ligand, and (d) the core-shell-ligand models. In these diagrams, differently shaded regions correspond to different materials with associated acoustic properties (as defined by $\rho$, $v_l$ and $v_t$).}
\label{fig:fig1}
\end{figure}

The Raman active modes of a homogenous elastic sphere are types of pure compressional modes, as given by
\begin{equation}
\mathbf{u}_{lm}(\mathbf{r},t) = \nabla \Phi_{lm}(\mathbf{r}),
\end{equation}
where
\begin{equation}
\Phi_{lm}(\mathbf{r}) = \left[A j_l\left(\frac{\omega_{lm}}{v_l}r \right) + By_l\left(\frac{\omega_{lm}}{v_l}r \right)\right]Y_l^m(\theta,\phi),
\end{equation}
and $Y_l^m(\theta,\phi)$ is the spherical harmonic function, $A$ and $B$ are constants, and $j_l$ and $y_l$ are spherical Bessel functions of the first and the second kinds, respectively. Among these compressional modes, only the radial breathing ($l=0$) and ellipsoidal ($l=2$) modes are Raman active~\cite{Duval1992} and are illustrated in Fig.~\ref{fig:fig1}a. In particular, the lowest frequency radial breathing mode ($l=0$ and $m=0$) has been the focus of many experimental investigations,~\cite{Cerullo1999, Son2006,Huxter2009,Chassaing2009} including our recent study of CdSe QDs.~\cite{Mork2016a,Mork2016b} In this manuscript, we focus specifically on modeling this radial breathing mode. Thus, for notational simplicity, we will denote the mode and its frequency without subscripts as $\mathbf{u}(\mathbf{r},t)$ and $\omega$, respectively.  

\subsection{$l=0$ mode in colloidal QDs} 
For the eigenmode corresponding to the lowest breathing mode, only the radial component of the displacement field is non-zero, which we denote as
\begin{equation}
u_r(r,t) = -q[A j_{1}(q r) + B y_{1}]\exp{(-i\omega t)},
\label{eq:2a}
\end{equation}
where $q\equiv \omega/v_l$. The strain tensor, $\bm{\varepsilon}=\frac{1}{2}[\nabla \mathbf{u} +(\nabla \mathbf{u})^T]$, associated with Eq.~\ref{eq:2a} has only three non-zero components, $\varepsilon_{rr} ={\partial u_r}/{\partial r}$ and $\varepsilon_{\theta \theta} = \varepsilon_{\phi \phi} = {u_r}/{ r}$. Based on the isotropic linear elastic stress-strain relation, the normal stress on the spherical surface is given by
\begin{eqnarray}
\bm{\sigma} \cdot \mathbf{\hat r} &=& \sigma_{rr}(r,t) \nonumber \\
&=& 2 \rho v_t^2 {\varepsilon_{rr}(r,t)} + \rho \left( v_l^2-2 v_t^2\right)\left[\varepsilon_{\theta \theta}(r,t)+\varepsilon_{\phi \phi}(r,t))\right] \nonumber\\
&=& -\frac{2 v_t^2 \rho}{r^2}  \bigg\{A \left[- \frac{Q^2 r^2}{2}j_0(qr) + 2 qrj_{1}(qr) \right]+ \nonumber\\
&B& \left[- \frac{Q^2 r^2}{2}y_0(qr) + 2 qry_{1}(qr) \right]\bigg\} \exp{(-i\omega t)} \label{eq:2b},
\end{eqnarray}
where $\bm{\sigma}$ is the stress tensor and $Q \equiv \omega / v_t$. Here constants $A$ and $B$ are determined in order to satisfy the boundary conditions.  

\subsection{The core-only model}
We model the low frequency acoustic properties of an isolated inorganic core with a \textit{core-only} model. In this model, the  stress and diplacement fields of a semiconductor nanocrystal are characterized simply by its radius, $R$, sound velocities, $v_l$ and $v_t$, and density, $\rho$. These properties, along with a description of the boundary conditions, combine to determine properties of the radial breathing mode of the nanocrystal. Two limiting cases are the free boundary, defined by the vanishing normal stress at the surface, $\sigma_{rr}(R,t)=0$,~\cite{Lamb1881} and the rigid boundary, defined by the vanishing displacement at the surface, $u_r(R,t) = 0$.~\cite{Saviot1996} In either case, this model predicts a Raman frequency shift that is given by $\nu = \frac{\omega}{2\pi c}= \alpha \frac{v_l}{2\pi c}\frac{1}{R}$, where $c$ is the speed of sound and $\alpha\equiv qR$ is a dimensionless frequency that depends on the details of the boundary condition. Specifically, the value of $\alpha$ is given by the characteristic root of the secular equation $f(\alpha) =0$, where
\begin{equation}
f(\alpha) = \tan(\alpha)-\alpha
\end{equation}
for the case of rigid boundary condition and
\begin{equation}
f(\alpha) = \tan(\alpha) - \frac{\alpha}{1-\frac{v_l^2}{4v_t^2}\alpha^2}
\end{equation}
for the case of free boundary condition.

\subsection{The core-ligand model}  
The \textit{core-ligand} model extends the previous model to include the effects of surface ligands on the vibrations of the inorganic core. Particularly, this model describes the vibrations of the QD core against a compressible ligand layer of finite thickness with a fixed outer boundary. We model the core (C) as a sphere of radius $R_\mathrm{C}$ that is nested within the center of spherical ligand shell (L) with radius $R_\mathrm{L}$ such that the thickness of the ligand shell is given by $d_\mathrm{L} = R_\mathrm{L} -R_\mathrm{C}$. As illustrated in Fig.~\ref{fig:fig1}c, the acoustic properties of the core (as specified by $\rho^{(\mathrm{C})}$, $v_l^{(\mathrm{C})}$, and $v_t^{(\mathrm{C})}$) are defined separately from that of the ligand layer (as specified by $\rho^{(\mathrm{L})}$, $v_l^{(\mathrm{L})}$, and $v_t^{(\mathrm{L})}$). We set a matching boundary condition at the core-ligand interface (\textit{i.e.}, at $r=R_\mathrm{C}$) and a rigid boundary condition at the outer ligand boundary (\textit{i.e.}, at $r=R_\mathrm{L}$). Formally, these boundary conditions are represented as
\begin{eqnarray}
u_r^{(\mathrm{C})}(r,t)&= u_r^{(\mathrm{L})}(r,t) & \mathrm{~at~}r = R_\mathrm{C}, \nonumber \\
\sigma_{rr}^{(\mathrm{C})}(r,t)&= \sigma_{rr}^{(\mathrm{L})}(r,t) & \mathrm{~at~}r = R_\mathrm{C}, \nonumber \\
u_r^{(\mathrm{L})}(r,t) &= 0~~~~~~~ & \mathrm{~at~}r = R_\mathrm{L}.
\label{eq:2}
\end{eqnarray}
Additionally, we impose a condition $B=0$ on $u_{r}^{(\mathrm{C})}$ and $\sigma_{rr}^{(\mathrm{C})}$ to prevent their divergence at the center of the core. 

\subsection{The core-shell-ligand model} 
We further extend the previous model to describe QDs with core-shell structure and surface ligands. As illustrated in Fig.~\ref{fig:fig1}d, this \textit{core-shell-ligand} model includes a spherical inorganic core (C) of radius $R_\mathrm{C}$, embedded within a spherical inorganic shell of radius $R_\mathrm{S}$. 
The inorganic core-shell structure is then embedded within a spherical ligand shell (L) of radius $R_\mathrm{L}$. The acoustic properties of each layer are defined separately as $\rho^{({j})}$, $v_l^{({j})}$, and $v_t^{({j})}$, where the superscript $j=$ C, S, or L, for the core, shell, or ligand materials, respectively. We assume all the layers in the onion-like structure vibrate as a single composite structure. The boundary conditions for this system are
\begin{eqnarray}
u_r^{(\mathrm{C})}(r,t)&= u_r^{(\mathrm{S})}(r,t) & \mathrm{~at~}r = R_\mathrm{C}, \nonumber \\
\sigma_{rr}^{(\mathrm{C})}(r,t)&= \sigma_{rr}^{(\mathrm{S})}(r,t) & \mathrm{~at~}r = R_\mathrm{C}, \nonumber \\
u_r^{(\mathrm{S})}(r,t)&= u_r^{(\mathrm{L})}(r,t) & \mathrm{~at~}r = R_\mathrm{S}, \nonumber \\
\sigma_{rr}^{(\mathrm{S})}(r,t)&= \sigma_{rr}^{(\mathrm{L})}(r,t) & \mathrm{~at~}r = R_\mathrm{S}, \nonumber \\
u_r^{(\mathrm{L})}(r,t) &= 0~~~~~~~ & \mathrm{~at~}r = R_\mathrm{L}.
\label{eq:3}
\end{eqnarray}
As in the core-ligand model, we constrain $B=0$ for $u_r^{(\mathrm{C})}$ and $\sigma_{rr}^{(\mathrm{C})}$. 

For this class of models (\textit{i.e.}, the core-ligand and the core-shell-ligand models), the set of boundary conditions gives rise to a system of linear equations that can be solved to compute the frequency $\omega$. In Appendices~\ref{sec:app1} and~\ref{sec:app2}, we describe the details of these systems of equations for the core-ligand and the core-shell-ligand models, respectively.

\section{Results}

\subsection{Size dependence of acoustic phonon energy}
\begin{figure}[h!] 
\centering
\includegraphics{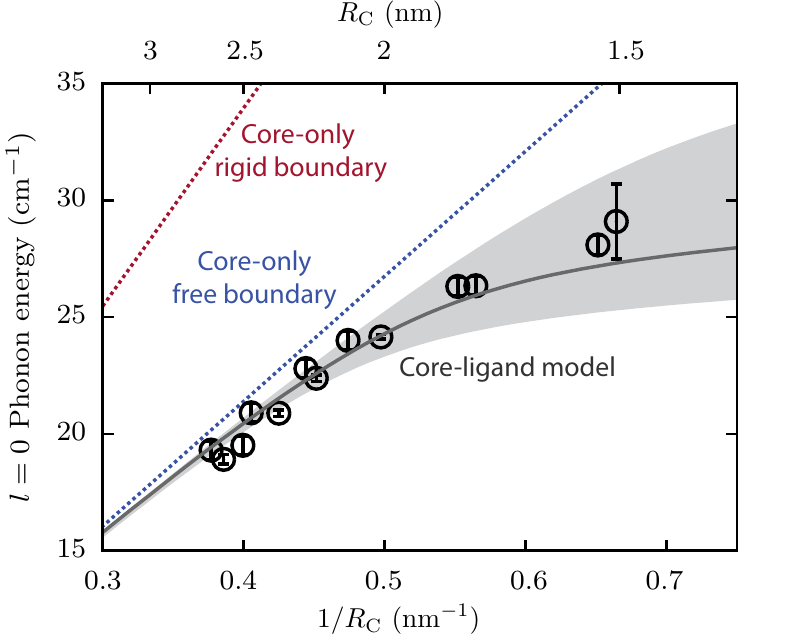}
\caption{The effect of QD size on the phonon energy of the radial breathing mode. Circles denote the experimental results with error bars corresponding to the standard deviation among at least four measurements. 10 cm$^{-1}$ = 0.3 THz.
Dotted lines indicate the prediction of the core-only model for rigid (red) and free (blue) boundary conditions.
The grey solid line is a fit to data using the core-ligand model, and the shaded region represents the uncertainty in the fit based on the uncertainty in QD size. (see Supplementary Material Section S2~\cite{Note1} for QD sizing information.)} 
\label{fig:fig2}
\end{figure}

We validate our model by comparing model predictions to Raman spectra taken on QD thin films. 
Specifically, we have synthesized thin films of CdSe QDs capped with octadecylphononic acid (ODPA) native ligands through a standard hot-injection method~\cite{Peng2001}.
We have collected the Stokes and anti-Stokes Raman spectra between 10 and 100 cm$^{-1}$ (about 0.3 to 3 THz) using a 785 nm narrow linewidth laser as the nonresonant Raman pump.
By varying the size of the QDs in each film, we are able to measure how acoustic vibrations depend on QD size.
The Raman spectra reveal a monotonic redshift in both the $l=0$ and $l=2$ phonon features with increasing nanoparticle size (see Fig. S7~\footnote{See Supplementary Material at [URL] for details of the nanocrystal synthesis and characterization, Raman spectral fitting, and various continuum modeling work discussed in this article.}), as has been observed previously.~\cite{Verma1999,Chassaing2009,Mork2016a,Mork2016b}
As Fig.~\ref{fig:fig2} illustrates, the scaling of vibrational energies with QD size is consistent with a core-only model for relatively large cores ($R\gtrsim2$ nm).

The core-only model predicts a linear relationship between $\nu$ and $1/R$, with a slope that depends on the details of the boundary conditions.
For comparison to experiment, we treat the CdSe core as a uniform material with the elastic parameters of bulk wurtzite CdSe: $\rho^{(\mathrm{C})}=5.81$ g/cm$^3$, $v_l^{(\mathrm{C})} = 3559$ m/s, and $v_t^{(\mathrm{C})} = 1607$ m/s.~\cite{Palmer2008,Murray_webpage,Madelung2004}
As Fig~\ref{fig:fig2} illustrates, we observe that for either choice of boundary conditions the core-only model tends to overpredict the value of the phonon energy for a given QD size.
Notably, the model with the free boundary condition is in much better agreement than that of the rigid boundary condition, suggesting that shape distortions are fundemental to the low frequency vibrations of QDs.

To model the effects of ligands on the vibrational energies of QDs, we apply the core-ligand model to the same system, retaining the same bulk CdSe wurtzite parameters to describe the QD core.
We assume that the ligands comprise a shell of constant thickness, $d_\mathrm{L}=R_\mathrm{L}-R_\mathrm{C}$, and uniform density, $\rho^{(\mathrm{L})}$, regardless of QD size.
We derive the ligand layer thickness to be $d_\mathrm{L}=1.3$ nm based on X-ray scattering measurements of the interparticle spacing of similar sized QDs capped with oleic acid (OA),~\cite{Weidman2014} which has similar hydrocarbon length to ODPA.
Specifically, we find that the size-dependent $l=0$ phonon energies of ODPA capped QDs synthesized by the standard hot-injection method and those measured for OA capped QDs synthesized by the seeded growth method have comparable energies. (see Fig. S8.~\cite{Note1})
Likewise, we assume that the density of the ligand shell is comparable to that of hydrocarbon-based organic polymers (\textit{e.g.}, polyethylene), which is about $\rho^{(\mathrm{L})} =0.9$ g/cm$^3$.~\cite{Selfridge1985}

After fixing $\rho^{(\mathrm{L})}$ and $d_\mathrm{L}$, we parameterize the longitudinal sound velocity of the ligand layer, $v_l^{(\mathrm{L})}$, by fitting the core-ligand model to experimental data. 
Notably, the use of a rigid boundary condition at the outer surface of the ligand layer eliminates any dependence of $\omega$ on the transverse ligand sound velocity, $v_t^{(\mathrm{L})}$.
We find that the fitted value of $v_l^{(\mathrm{L})} = 2424 \pm 70$ m/s for the OPDA sound velocity is in remarkably good agreement with the experimentally measured value of 2430 m/s for high-density polyethylene,~\cite{Selfridge1985} a similarly saturated hydrocarbon material.
Additional details on the data fitting procedure and ligand parameter estimations used in the core-ligand model are described in Supplementary Material Section S6.~\cite{Note1}


\subsection{Sound velocities of QD surface alkanethiol ligands}
The low frequency vibrations of QDs depend on the identity of ligands that passivate their surfaces.
We analyze this dependence by applying our model to a series of measurements carried out on films comprised of CdSe cores with fixed size but with different ligands.
Here we consider the case where ligand exchange has been carried out to generate CdSe QDs with straight-chain alkanethiols of variable length, ranging from hexanethiol (C6) to octadecanethiol (C18).~\cite{Mork2016a}
We assume that the ligand density is constant, $\rho_\mathrm{L}=0.9 \mathrm{g/cm^3}$, for all lengths of alkanethiol ligands.
We assign ligand layer thickness, $d_\mathrm{L}$ based on X-ray scattering studies of similar sized PbS nanocrystals,~\cite{Weidman2014} whose values are tabulated in Table~\ref{tab:t1}. The only fitting parameter is, therefore, the longitudinal sound velocity of the ligand layer, $v_l^{(\mathrm{L})}$. 

\begin{figure}[h!] 
\centering
\includegraphics{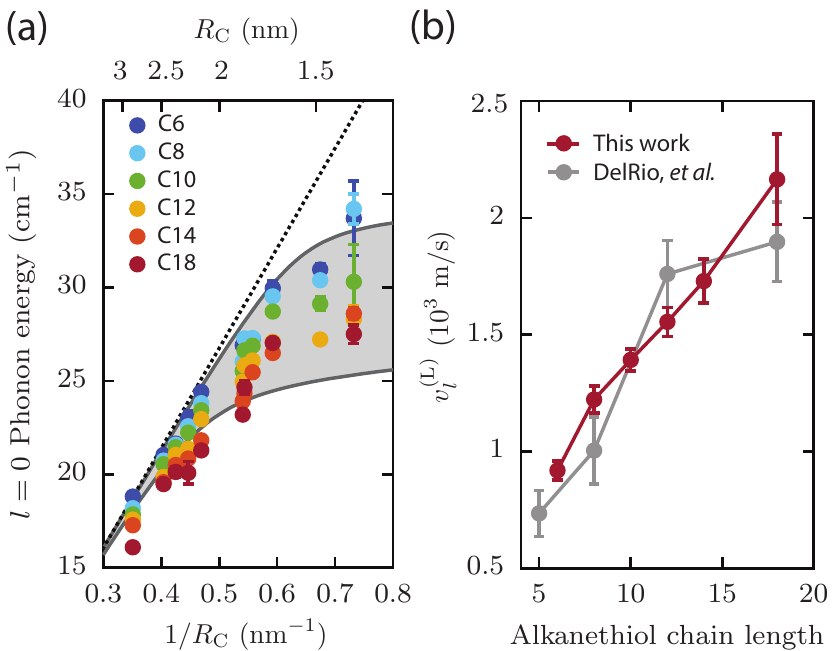}
\caption{The effect of varying ligand length on the acoustic vibrations of QDs. (a) Ligand length dependence of the $l=0$ QD phonon energy. The dotted line represents the prediction based on the core-only model with a free-boundary condition. Colored circles are the experimentally measured phonon energies for alkanethiol ligands with carbon number indicated in the legend.~\cite{Mork2016a} The grey shaded region represents the predictions from the core-ligand model, which is bounded anove and below by the fits to the hexanethiol (C6) and octadecanthiol (C18) data, respectively. (b) Fitted ligand sound velocity as a function of alkanethiol chain length (number of carbons) from the QD data in (a) compared to the measured values for monolayers of the same molecules self-assembled on planar gold surfaces.~\cite{DelRio2009} Error bars denote one standard deviation uncertainty in the corresponding sound velocity for both measurements~\citep{DelRio2009} and fits.}
\label{fig:fig3}.  
\end{figure}

Fig.~\ref{fig:fig3}a shows that the core-ligand model captures most of the observed variation in acoustic mode energy with changing ligand length. 
The gray lines in this figure delineate the core-ligand model fits to the hexanethiol (C6 in Fig.~\ref{fig:fig3}a) and octadecanthiol (C18 in Fig.~\ref{fig:fig3}a) data series. 
For this data, the fitted ligand sound velocities compare favorably to earlier mechanical measurements by DelRio \textit{et al.}~\cite{DelRio2009} for linear alkanethiols self-assembled on gold surfaces. (Fig.~\ref{fig:fig3}b; see Supplementary Material Section S6.~\cite{Note1})
The close agreement between the sound velocities extracted from our model and previously measured elastic properties of surface-bound molecules in self-assembled monolayers (SAMs) implies a similarity in the elastic behavior of organic molecules bound in either flat (SAM) or curved (QD) geometries. 

Furthermore, we relate $v_l^{(\mathrm{L})}$ to the elastic stiffness tensor, $\mathbf{C}$, of the ligand shell, which is an analog of the spring constant in Hooke's law for multidimensional systems. In our model, the elastic stiffness moduli $C_{11}$ and $C_{12}$ relate the normal stress, $\sigma_{rr}$, to the stress components by $\sigma_{rr} = C_{11}\varepsilon_{rr} + C_{12}(\varepsilon_{\theta \theta} + \varepsilon_{\phi \phi})$. In particular, for an isotropic material, $C_{11}$ is the elastic stiffness modulus that describes the compression along the longitudinal axis and is related to the sound velocity by $C_{11} = v_l^2 \rho$. 
Thus, variations in $v_l^{(\mathrm{L})}$ with respect to the ligand length imply that the elastic stiffness modulus also depends on the ligand length. Our computed values of $C_{11}^{(\mathrm{L})}$ are listed in Table~\ref{tab:t1}. Calculated stiffness moduli of various ligands ($C_{11}^{(\mathrm{L})}=0.8$--5 GPa) are comparable to values that have been estimated for other colloidal nanocrystal assemblies using atomic force microscopy (AFM)~\cite{Podsiadlo2010} as well as those of self-assembled monolayers on gold substrates,~\cite{DelRio2009} and organic polymers.~\cite{Selfridge1985}

\begin{table}
\caption{Surface ligand parameters\label{tab:t1} }
\begin{ruledtabular}
\begin{tabular}{l|l|l|l}
ligand& $d_\mathrm{L}$ (nm)\footnote{Estimated based on half of the measured interparticle spacing of similar sized PbS QDs with corresponding ligands derived from a X-ray scattering study.~\cite{Weidman2014} See Supplementary Material Section S6.~\cite{Note1}} & $v_l^{(\mathrm{L})}$ (m/s)\footnote{Calculated by fitting the core-ligand model to the size-dependent $l=0$ phonon energies of CdSe QDs measured by Raman spectroscopy, assuming ligand layer density of $\rho^{(\mathrm{L})} = 0.9$ g/cm$^3$. Error represents one standard deviation in the parameter uncertainty.} & $C_{11}^{(\mathrm{L})}$ (GPa)\footnote{Calculated by $C_{11}^{(\mathrm{L})} = [v_l^{(\mathrm{L})}]^2 \rho^{(\mathrm{L})}$.} \\ \hline
octadecylphononic acid & 1.3 & $2424 \pm 70$ & $5.28 \pm 0.31$ \\
oleic acid  & 1.3 & $2439 \pm 253$ & $5.35 \pm 1.11$\\
hexanethiol (C6) & 0.45 & $919 \pm 42$ & $0.76 \pm 0.07$ \\
octanethiol (C8) & 0.61 & $1222 \pm 59$ & $1.34 \pm 0.13$\\
decanethiol (C10) & 0.76 & $1392 \pm 48$ & $1.74 \pm 0.12$ \\
dodecanethiol (C12) & 0.91 & $1554 \pm 63$ & $2.17 \pm 0.18$\\
tetradecanethiol (C14)  & 1.07 & $1730 \pm 95$ & $2.69 \pm 0.30$ \\
octadecanethiol (C18)& 1.38 & $2165 \pm 197$ & $4.22 \pm 0.78$ \\ 
\end{tabular}
\end{ruledtabular} 
\end{table}

\subsection{Acoustic phonons in core-shell QDs}
\begin{figure}[h!] 
\centering
\includegraphics{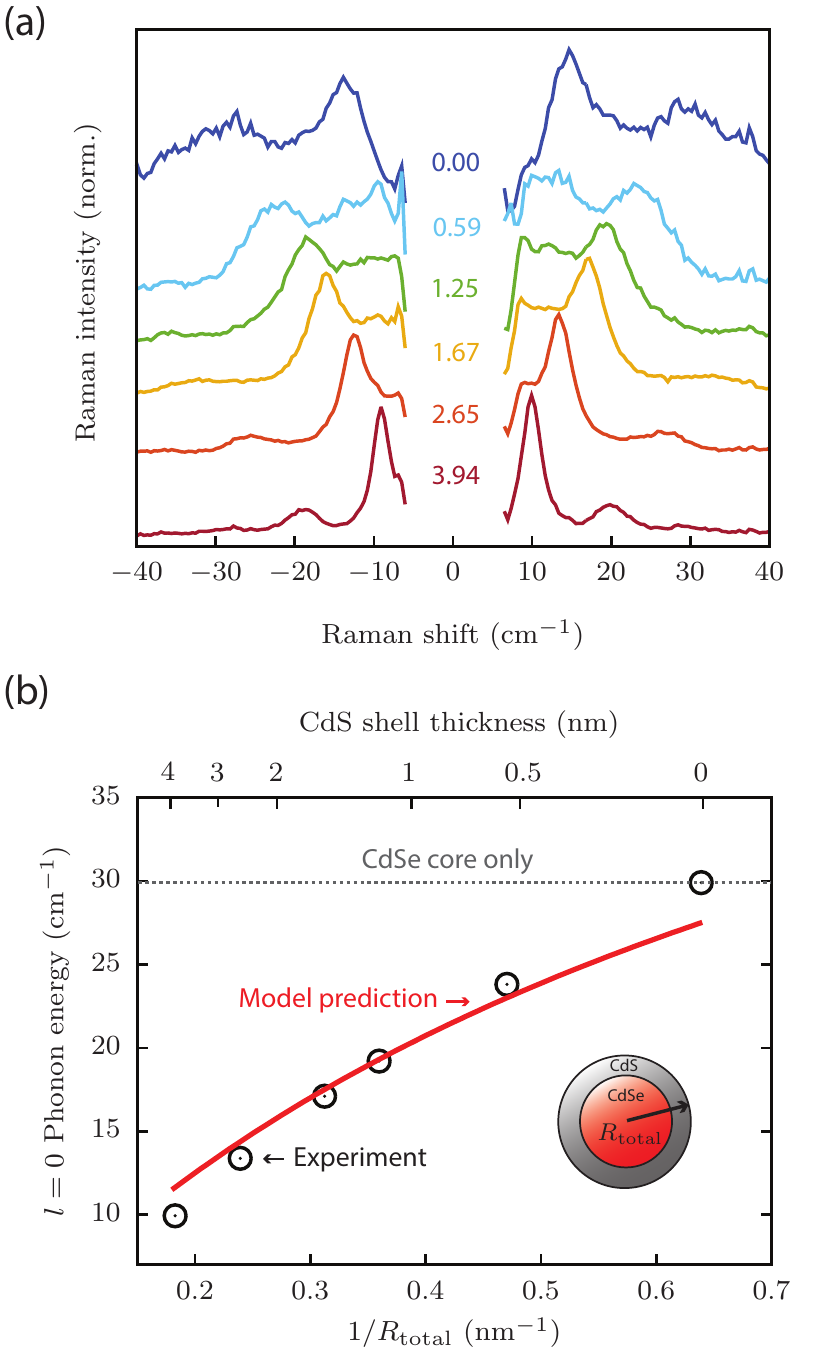}
\caption{Phonon energies of core-shell QD heterostructures. (a) Raman spectra of a series of CdSe/CdS core/shell QDs with identical core size but increasing CdS shell thickness from top (blue line) to bottom (red line), with labels in the middle indicating the measured thickness (in nm) of the CdS shell. (b) Size-dependence of the $l=0$ phonon energy corresponding to the spectra shown in (a). Open circles denote experimental data, and solid line indicates the prediction of the core-shell-ligand model.}
\label{fig:fig4}
\end{figure}

The highly luminescent QD materials used in light-emitting applications nearly always include an epitaxially grown shell of a higher band-gap semiconductor around the central core.~\cite{Kagan2016} To investigate how this inorganic shell impacts the acoustic phonon spectrum of core-shell heteronanocrystals, we have synthesized a set of nanocrystals with a constant size CdSe core and variable thickness CdS shell. (see Supplementary Material Section S3.~\cite{Note1}) The low-frequency Raman spectra of these particles reveal a monotonic redshift in acoustic phonon frequencies with increasing shell thickness (Fig.~\ref{fig:fig4}a). We also observe an additional peak at about twice the frequency of the $l=0$ peak, which has been identified through our previous temperature dependent measurements as a higher spherical harmonic---the $n = 2$, $l = 0$ eigenmode---rather than a two-phonon scattering event.~\cite{Mork2016b}

We model these experiments with the core-shell-ligand model, using parameters for bulk CdS (wurtzite CdS bulk parameters: $\rho^{(\mathrm{S})}$ = 4.82 g/cm$^3$, $v_l^{(\mathrm{S})} = 4289$ m/s, and $v_t^{(\mathrm{S})} = 1870$ m/s)~\cite{Palmer2008,Murray_webpage,Madelung2004} to describe the properties of the inorganic shell. 
We find that without any further fitting of the ligand parameters, the model agrees well with the experiment. The ability to transfer the ligand parameters across different models suggests that the nature of the chemical linkage at the core/ligand interface, \textit{i.e.}, ligand-CdSe versus ligand-CdS, does not significantly affect the contribution of the ligands to low frequency vibrational modes. 

\section{Discussion}
The observation that the low frequency vibrational modes of small QDs deviate from the $1/R$ scaling of Lamb's model has been interpreted to indicate a breakdown of the assumptions that underlie traditional elastic continuum theories. 
It has thus been suspected that small QDs (\textit{i.e.}, with radius $R\sim 2$ nm) are not well approximated as uniform elastic spheres, due to, for instance, microscopic effects such as surface-specific lattice restructuring and collective ligand dynamics. 
Our model highlights that these types of effects are unnecessary to explain the small QD scaling deviations. In fact, these deviations are entirely consistent with the framework of elastic continuum theory.
As we have demonstrated, the failure of Lamb's model is to omit the elastic effects of the ligands that passivate the surface of the semiconductor nanocrystal.
These effects are negligible when the mass of the inorganic core is much larger than that of the ligand shell, but they become increasingly significant as the size of the nanoscrystal is reduced and becomes comparable to that of the ligand shell thickness.

Our model implies a physical description of QD films in which an array deformable inorganic nanocrystals is embedded within a uniform ligand matrix.
This description emerges primarily from the requirement of rigid boundary conditions at the exterior of the ligand layer. 
As illustrated in Supplementary Material, the use of free ligand boundary conditions in our model yields  breathing mode frequencies that are significantly lower than those from experimental measurements. (see Fig. S9.~\cite{Note1})
This suggests that in QD films the ligand shell volume does not undergo significant shape distortions but, rather, is constrained by the surrounding environment. 
This is more consistent with a model in which adjacent ligand shells solublize each other to form a partially inter-digitated background ligand environment.~\cite{Goodfellow2015,Boles2015} 

\begin{figure}[h!] 
\centering
\includegraphics{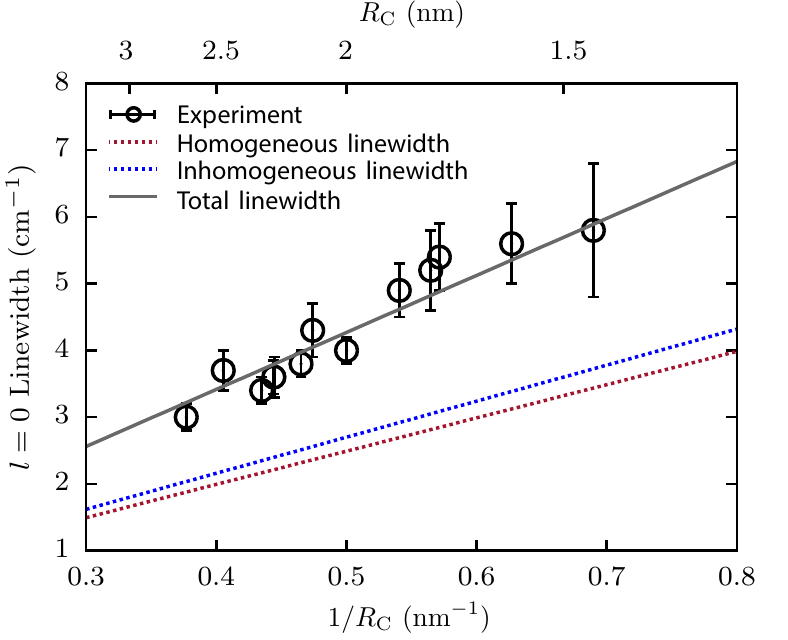}
\caption{Phonon linewidths of CdSe QDs. Full-width at half-maximum (FWHM) of Gaussian fit to the $l=0$ phonon peak for two size series of CdSe QDs with native phosphonic acid ligands (circles) compared to the predicted homogenous and the inhomogeneous linewidth (assuming 10 \% polydispersity) based on the dissipative medium model (Appendix C).}
\label{fig:fig5}.  
\end{figure}

Further details about the nature of the background ligand environment can be derived by analyzing the lineshape of the lowest frequency Raman mode.
There are many possible contributions to this lineshape, including QD size polydispersity, multiphonon effects, and anharmonic coupling at the QD surface.~\cite{Alivisatos1996,LiuJ2015} 
Raman spectra of core-only dots whose phonon energies have been plotted in Fig.~\ref{fig:fig2} (see Fig. S7~\cite{Note1} for the corresponding Raman spectra) reveal that the linewidth includes contributions that vary with CdSe core size. 
In particular, Fig.~\ref{fig:fig5} reveals that the lowest frequency peak narrows with increasing core size in a manner that scales inversely with the radius of the core.
This scaling suggests that the spectral narrowing may be due to ligands exerting an acoustic impedance effect.
Such an effect has been observed previously for nanoparticles embedded in glass matrix.~\cite{Verma1999,Saviot2004}
Unfortunately, our purely elastic model lacks any information about spectral lineshapes.
To address this effect, we thus modify our model to describe the ligand layer as a dissipative medium rather than a mass-loading boundary layer.
As described in more detail in Appendix~\ref{sec:app3}, this dissipative ligand model damps the acoustic vibration of the elastic core.
This model yields a finite (homogeneous) linewidth that can be compared directly to the experimental results plotted in Fig.~\ref{fig:fig5}.

As illustrated in Fig.~\ref{fig:fig5}, the dissipative ligand model predicts approximately the correct scaling for linewidth, but underpredicts its value by about a factor of two.
We attribute this apparent discrepancy to the presence of inhomogeneous broadening (due to variations in QD size/shape), which is not included in our model.
We note that previous measurements in our lab have shown that for a given QD thin film, the linewidth is approximately invariant to temperature over the range 77-300 K.~\cite{Mork2016b}
These experiments allow us to estimate the contribution to the $l=0$ linewidth from inhomogeneous broadening as a function of QD size.
Based on this, we assume a Gaussian inhomogeneous broadening arising from 10\% polydispersity, which yields predicted total linewidths that are consistent with experimental observations.

\section{Conclusions}
In this manuscript, we have demonstrated that the anomalous scaling of low-frequency vibrational energies that have been observed for small QDs is consistent with a simple elastic continuum description of the nanocrystal.
Here we have extended a traditional elastic continuum model (Lamb's model) to include the elastic response of the ligand layer that passivates the nanocrystal surface. 
We have highlighted that the effects of this ligand layer are negligible for larger QDs but play an important role in tuning the acoustic properties of QDs that are smaller than about 2 nm in radius.
We have demonstrated that our extended elastic continuum model is in quantitative agreement with Raman measurements carried out on a variety of QDs with different sizes, surface ligands, and compositions.

By comparing to experiments, we have shown that our elastic continuum model can be used to estimate the acoustic properties of the ligand layer. 
In fact, we have found that the sound velocity, $v_l^{(\mathrm{L})}$, of the ligand layer that surrounds QDs is nearly identical to that of analogous bulk organic polymer systems and self-assembled monolayers.
This finding implies that the elastic properties of a given type of ligands bonded to nanostructures could be universal, enabling to transfer ligand material parameters across various materials in both bulk-phase and nanocomposites.

\section*{Author contributions}
E.M.Y.L. developed the theoretical model and A.J.M. performed experiments. All authors discussed results and contributed in writing the manuscript. 

\begin{acknowledgments}
This work was supported by the U.S. National Science Foundation Division of Chemistry and Division of Materials Research under Award Number 1452857. E.M.Y.L. and A.J.M. have been partially supported by the U.S. National Science Foundation Graduate Research Fellowship Program under Grant No. 1122374. 
\end{acknowledgments}

\appendix
\section{\label{sec:app1}Matrix form of the core-ligand model}
The set of boundary conditions in Eq.~\ref{eq:2} yields a system of linear equations for the constants $A_{\mathrm{C}}$, $A_{\mathrm{L}}$, and $B_{\mathrm{L}}$ in the form of $\mathbf{Mx=0}$, where
\begin{equation}
\x = \left[ \begin{array}{c}
A_{\mathrm{C}}\\
A_{\mathrm{L}}\\
B_{\mathrm{L}}
\end{array} \right],
\end{equation}
and $\mathbf{M}$ is a 3x3 matrix whose non-zero elements $M_{ij}$'s are
\begin{eqnarray}
M_{11} &=& \alpha_{\mathrm{CC}}^2j_1(\alpha_{\mathrm{CC}}),\\
M_{12} &=& -\frac{v_{l}^{(\mathrm{C})}}{v_{l}^{(\mathrm{L})}} \alpha_{\mathrm{CC}}^2j_1(\alpha_{\mathrm{CC}}),\\
M_{13} &=& -\frac{v_{l}^{(\mathrm{C})}}{v_{l}^{(\mathrm{L})}} \alpha_{\mathrm{CC}}^2y_1(\alpha_{\mathrm{CC}}),\\
M_{21} &=& -\frac{\rho^{(\mathrm{C})}}{\rho^{(\mathrm{L})}}\alpha_{\mathrm{CC}}^2 \times \nonumber \\
&&\left\lbrace \alpha_{\mathrm{CC}} j_0(\alpha_{\mathrm{CC}}) -4\left[\frac{v_{t}^{(\mathrm{C})}}{v_{l}^{(\mathrm{C})}}\right]^2j_1(\alpha_{\mathrm{CC}})\right\rbrace, \\
M_{22} &=& \alpha_{\mathrm{CC}}^2\left\lbrace\alpha_{\mathrm{CC}} j_0(\alpha_{\mathrm{LC}}) -\frac{4 [v_{t}^{(\mathrm{L})}]^2}{v_{l}^{(\mathrm{C})}v_{l}^{(\mathrm{L})}}j_1(\alpha_{\mathrm{LC}})\right\rbrace,\\
M_{23} &=& \alpha_{\mathrm{CC}}^2\left\lbrace\alpha_{\mathrm{CC}} y_0(\alpha_{\mathrm{LC}}) -\frac{4[v_{t}^{(\mathrm{L})}]^2}{v_{l}^{(\mathrm{C})}v_{l}^{(\mathrm{L})}}y_1(\alpha_{\mathrm{LC}})\right\rbrace,\\
M_{32} &=& \alpha_{\mathrm{LL}}^2j_1(\alpha_{\mathrm{LL}}),\\
M_{33} &=& \alpha_{\mathrm{LL}}^2y_1(\alpha_{\mathrm{LL}}),
\end{eqnarray}
where $\alpha_{ij} = q_iR_j$ and $q_i = \omega/v_{l}^{(i)}$ for $i,j$ denoting the type of material, \textit{e.g.}, (C)=core and (L)=ligand. 

\section{\label{sec:app2}Matrix form of the core-shell-ligand model}
Having an additional inorganic shell layer compared to the core-ligand model leads to a total of five unknown constants, which are again determined by solving the equation $\mathbf{Mx=0}$, where
\begin{equation}
\x = \left[ \begin{array}{c}
A_{\mathrm{C}}\\
A_{\mathrm{S}}\\
B_{\mathrm{S}}\\
A_{\mathrm{L}}\\
B_{\mathrm{L}}
\end{array} \right],
\end{equation}
and $\mathbf{M}$ is a 5x5 matrix whose non-zero elements $M_{ij}$'s are
\begin{eqnarray}
M_{11} &=& \alpha_{\mathrm{CC}}^2j_1(\alpha_{\mathrm{CC}}),\\
M_{12} &=& -\frac{v_{l}^{(\mathrm{C})}}{v_{l}^{(\mathrm{S})}} \alpha_{\mathrm{CC}}^2j_1(\alpha_{\mathrm{CC}}),\\
M_{13} &=& -\frac{v_{l}^{(\mathrm{C})}}{v_{l}^{(\mathrm{S})}} \alpha_{\mathrm{CC}}^2y_1(\alpha_{\mathrm{CC}}),\\
M_{21} &=& -\frac{\rho^{(\mathrm{C})}}{\rho^{(\mathrm{S})}}\alpha_{\mathrm{CC}}^2 \times \nonumber \\
&&\left\lbrace \alpha_{\mathrm{CC}} j_0(\alpha_{\mathrm{CC}}) -4\left[\frac{v_{t}^{(\mathrm{C})}}{v_{l}^{(\mathrm{C})}}\right]^2j_1(\alpha_{\mathrm{CC}})\right\rbrace, \\
M_{22} &=& \alpha_{\mathrm{CC}}^2\left\lbrace\alpha_{\mathrm{CC}} j_0(\alpha_{\mathrm{SC}}) -\frac{4 [v_{t}^{(\mathrm{S})}]^2}{v_{l}^{(\mathrm{C})}v_{l}^{(\mathrm{S})}}j_1(\alpha_{\mathrm{SC}})\right\rbrace\\
M_{23} &=& \alpha_{\mathrm{CC}}^2\left\lbrace\alpha_{\mathrm{CC}} y_0(\alpha_{\mathrm{SC}}) -\frac{4[v_{t}^{(\mathrm{S})}]^2}{v_{l}^{(\mathrm{C})}v_{l}^{(\mathrm{S})}}y_1(\alpha_{\mathrm{SC}})\right\rbrace,\\
M_{32} &=& \alpha_{\mathrm{SS}}^2j_1(\alpha_{\mathrm{SS}}),\\
M_{33} &=& -\frac{v_{l}^{(\mathrm{S})}}{v_{l}^{(\mathrm{L})}}\alpha_{\mathrm{SS}}^2 j_1(\alpha_{\mathrm{LS}}),\\
M_{34} &=& -\frac{v_{l}^{(\mathrm{S})}}{v_{l}^{(\mathrm{L})}}\alpha_{\mathrm{SS}}^2 y_1(\alpha_{\mathrm{LS}}),\\
M_{42} &=& -\frac{\rho^{(\mathrm{S})}}{\rho^{(\mathrm{L})}}\alpha_{\mathrm{SS}}^2 \times \nonumber \\
&&\left\lbrace\alpha_{\mathrm{SS}} j_0(\alpha_{\mathrm{SS}}) -4 \left[\frac{v_{t}^{(\mathrm{S})}}{v_{l}^{(\mathrm{S})}}\right]^2j_1(\alpha_{\mathrm{SS}})\right\rbrace,\\
M_{43} &=& \alpha_{\mathrm{SS}}^2\left\lbrace\alpha_{\mathrm{SS}} j_0(\alpha_{\mathrm{LS}}) -\frac{4[v_{t}^{(\mathrm{L})}]^2}{v_{l}^{(\mathrm{S})}v_{l}^{(\mathrm{L})}}j_1(\alpha_{\mathrm{LS}})\right\rbrace,\\
M_{44} &=& \alpha_{\mathrm{SS}}^2\left\lbrace\alpha_{\mathrm{SS}} y_0(\alpha_{\mathrm{LS}}) -\frac{4[v_{t}^{(\mathrm{L})}]^2}{v_{l}^{(\mathrm{S})}v_{l}^{(\mathrm{L})}}y_1(\alpha_{\mathrm{LS}})\right\rbrace,\\
M_{54} &=& \alpha_{\mathrm{LL}}^2j_1(\alpha_{\mathrm{LL}}),\\
M_{55} &=& \alpha_{\mathrm{LL}}^2y_1(\alpha_{\mathrm{LL}}),
\end{eqnarray}
where (S)=shell. 

\section{\label{sec:app3}The dissipative medium model}
Treatment of organic ligands as a dissipative medium that damps vibrational modes originating from an inorganic core (\textit{e.g.}, CdSe) with radius $R$ allows sounds waves to be transmitted to the surrounding medium ($r>R$) but not reflect back toward the core. Thus for $r>R$, we replace spherical Bessel functions with spherical Hankel function of the first kind, $h_n^{(1)}$ in Eqs.~\ref{eq:2a} and \ref{eq:2b} such that
\begin{eqnarray}
u_r(r,t) &=& -\frac{1}{r}q r A h_{1}(q r)\exp{(-i\omega t)},\\
\sigma_{rr}(r,t) &=& \frac{2 v_t^2 \rho}{r^2}  A \left[- \frac{Q^2 r^2}{2}h_0(qr) + 2 qrh_{1}(qr) \right]
\\ 
&\times& \exp{(-i\omega t)}.
\nonumber
\end{eqnarray}

With continuous stress and displacement boundary conditions at $r=R$, we solve for the non-linear system of equations, $\mathbf{f(x)}=0$, to calculate the complex-valued frequency, $\omega = \omega_1 + i\omega_2$,
\begin{equation}
\mathbf{f(x)} = \left[ \begin{array}{c}
\text{Re} (\det \mathbf{M})\\
\text{Im} (\det \mathbf{M})\\
\end{array} \right],
\end{equation}
where
\begin{equation}
\mathbf{x} = \left[ \begin{array}{c}
\omega_1\\
\omega_2 \\
\end{array} \right].
\end{equation}
$\mathbf{M}$ is a 2x2 matrix with elements,
\begin{eqnarray}
M_{11} &=& \alpha_{\mathrm{CC}}^2 j_1 (\alpha_{\mathrm{CC}}) \\
M_{12} &=& -\frac{v_{l}^{(\mathrm{C})}}{v_{l}^{(\mathrm{L})}}\alpha_{\mathrm{CC}}^2 h_1^{(1)}(\alpha_{\mathrm{LC}})\\
M_{21} &=& \frac{\rho^{(\mathrm{C})}}{\rho^{(\mathrm{L})}}\alpha_{\mathrm{CC}}^2\times \nonumber \\
&&\left\lbrace 4j_1(\alpha_{\mathrm{CC}}) -\alpha_{\mathrm{CC}} \left[\frac{v_{l}^{(\mathrm{C})}}{v_{l}^{(\mathrm{L})}}\right]^2 j_0(\alpha_{\mathrm{CC}}) \right\rbrace \\
M_{22} &=& \alpha_{\mathrm{CC}}^3 \left[ \frac{v_{l}^{(\mathrm{C})}}{v_{t}^{(\mathrm{C})}} \right]^2 h_0^{(1)}(\alpha_{\mathrm{LC}}) \nonumber\\
&& -4 \alpha_{\mathrm{CC}}^2 \frac{v_{l}^{(\mathrm{C})}}{v_{l}^{(\mathrm{L})}} \left[\frac{v_{t}^{(\mathrm{L})}}{v_{t}^{(\mathrm{C})}} \right]^2 h_1^{(1)}(\alpha_{\mathrm{LC}}).
\end{eqnarray}
The real part, $\omega_1$, is the phonon energy, and the imaginary part, $\omega_2$, is the phonon half-width at half-maximum of a Lorentzian homogenous broadening.~\cite{Verma1999} Using the octadecylphosphonic acid (ODPA) ligand parameters ($v_{l}^{(\mathrm{L})} = 2424$ m/s and $\rho^{(\mathrm{L})} = 0.9$ cm$^3$/g) and assuming ${v_{t}^{(\mathrm{L})}}/{v_{t}^{(\mathrm{C})}} \approx 0$, we find that the phonon energy is the same as that of the free-boundary core-only model (Fig.~\ref{fig:fig2}) while the homogenous linewidth is proportional to $1/R$ as shown in Fig.~\ref{fig:fig5}.

\bibliography{ref}

\end{document}